\documentclass[twocolumn,showpacs,superscriptaddress,pra,floatfix,footnote]{revtex4}
\usepackage{amsmath}
\usepackage{amssymb}
\usepackage{latexsym}
\usepackage{array}
\usepackage{amsfonts}
\usepackage{mathrsfs}
\usepackage{color}
\usepackage{hyperref}
\usepackage{graphicx}

\def\be{\begin{equation}}
\def\ee{\end{equation}}
\def\bea{\begin{eqnarray}}
\def\eea{\end{eqnarray}}
\def\ben{\begin{equation*}}
\def\een{\end{equation*}}
\def\bean{\begin{eqnarray*}}
\def\eean{\end{eqnarray*}}
\def\bma{\begin{mathletters}}
\def\ema{\end{mathletters}}
\def\bi{\begin{itemize}}
\def\ei{\end{itemize}}
\begin{document}

\title{Device and semi-device independent random numbers based on non-inequality paradox}

\author{Hong-Wei Li}
\affiliation{Key Laboratory of Quantum Information,University of Science and Technology of China,Hefei, 230026,
 China}
 \affiliation{Zhengzhou Information Science and Technology Institute, Zhengzhou, 450004, China}
\author{Marcin Paw{\l}owski}
\email{maymp@bristol.ac.uk}\affiliation{Institute of Theoretical Physics and Astrophysics, University of Gdansk, 80-952 Gdansk, Poland}
\affiliation{Department of Mathematics, University of Bristol, Bristol BS8 1TW, United Kingdom}
\author{Ramij Rahaman}\email{ramijrahaman@gmail.com}
\affiliation{Department of Mathematics, University of Allahabad, Allahabad-211002, U.P., India}
\author{Guang-Can Guo}
\affiliation{Key Laboratory of Quantum Information,University of Science and Technology of China,Hefei, 230026,
 China}
\author{Zheng-Fu Han}
\email{zfhan@ustc.edu.cn}\affiliation{Key Laboratory of Quantum Information,University of Science and Technology of China,Hefei, 230026,
 China}

 \date{\today}
\begin{abstract}

In this work, we propose device independent true random numbers
generation protocols based on non-inequality paradoxes such as Hardy's and
Cabello's non-locality argument. The efficiency of generating randomness in our protocols are far better than any other proposed
protocols certified by CHSH inequality or other non-locality test involving inequalities. Thus, highlighting non-inequality paradox as an important
resource for device independent quantum information processing in particular generating true randomness. As a
byproduct, we find that the non-local bound of the Cabello's argument
with arbitrary dimension is the same as the one achieved in the qubits system. More interestingly, we propose a new dimension
witness paradox based on the Cabello's argument, which can be used for
constructing semi-device-independent true random numbers generation
protocol.

\end{abstract}
\maketitle

\section{ Introduction}
Randomness is an important basic feature of nature, which has wide
applications in information processing including other prominent
fields {\em e.g.}, biology, chemistry, social science, {\em etc.}.
At present, the protocol behind most of the true random number
generators is based primarily on classical laws of physics and
fundamentally they are all deterministic in some underline theory.
Therefore, people are looking for new protocols which can produce
genuine private randomness. On the other hand, one of the intrinsic
feature of quantum mechanics is that it is a probabilistic theory
inherently. This probabilistic feature of quantum theory does not
come from the subjective ignorance about the pre-assigned value of a
dynamical variable in a quantum state, rather it represents the
probabilistic nature of finding a particular value of a dynamical
variable if that dynamical variable is measured. However, in any
real experiment the randomness of measurement outcomes of quantum
systems is unavoidably mixed-up with an apparent randomness that
results from noise or lack of control on the quantum devices.
Therefore, it is not so easy to generate ideal randomness even with
the assist of quantum technology and several attempts have been made
in this regards. To generate the certified quantum random number,
Colbeck \cite{Colbeck 1,Colbeck 2} introduced a {\em device
independent true random number generation (DITRNG)} protocol based
on the GHZ test \cite{GHZ}, while Pironio {\it et.al.} \cite{Pironio
1} proposed another DITRNG protocol based on the violation of CHSH
inequality \cite{CHSH}.

 In CHSH inequality based DITRNG protocol, players distrust the concerning bipartite entangled state and all the measurement devices as they might have
 been fabricated by a spiteful party Eve,
 the randomness can be guaranteed if the CHSH expression violate
the {\em local hidden variable(LHV)} bound. The maximal min-entropy
of the generated random number is about 1.23 and it occurs when the
corresponding CHSH expression attends its maximum quantum bound
$2\sqrt{2}$. Unfortunately, all the existing DITRNG protocols
require entanglement, which has the negative impact on the
complexity of the devices and the rate of the random number generation.
To get higher key rate recently, Li {\em et al.} \cite{Li1,Li2,Li3}
proposed a {\em semi-device independent true random number
generation (SDITRNG)} protocol, where the true randomness certified if the corresponding dimension witness
inequality \cite{Witness,Witness 1,Witness 2,Witness 3} violates
its LHV bound.

In general, the device independent scenarios have also been proposed
in the context of several other information processing tasks, for
example, quantum cryptography\cite{QC}, state estimation \cite{SE}
{\em etc.}. However, all of these protocols are mostly based
either on CHSH inequality \cite{CHSH}, Mermin's
inequality \cite{MERMIN} or Chained inequality \cite{CHAIN1,CHAIN2}.
Here we propose a DITRNG protocol based on some other non-locality
tests without involving any statistical inequalities such as Hardy's
paradox \cite{Hardy} and Cabello's paradox \cite{Cabello}. The two
paradoxes respectively shows a direct contradiction between quantum
theory and {\em local-realistic (LR) theory} with the help of two-qubit correlations. We find that our DITRNG
protocols based on Hardy paradox or Cabello paradox are more
efficient than other proposed inequality based DITRNG protocols in
various context.

We start with a description of a non-locality test without using any statistical inequality known as Hardy paradox, which will allow us to
 formulate the DITRNG protocol based on non-inequality paradox. The maximal bound of generated randomness in this case is larger than the CHSH
 based DITRNG protocol. Next we present the protocol based on another non-inequality test called Cabello paradox and show that the efficiency of this is even better than Hardy paradox.
Finally, we propose a new dimension
witness paradox without involving any inequality, which also allow us to generate much more random number compare to other proposed SDITRNG protocols.
All in all, our result shows that there is a practical advantage of using non-inequality
paradox based protocol to generate true randomness over the CHSH inequality based protocols.

\section{ Non-local paradox without inequality}
\subsection{Hardy paradox} In 1992, Hardy \cite{Hardy} suggested an non-locality test without using any statistical inequality for two two-level system in
comparison with Bell-CHSH test. Consequently, Hardy's test known as
an non-locality test without inequality. Consider a physical system
shared between two remote parties say, Alice and Bob. Let Alice can
perform measurement on her subsystem chosen randomly from two
positive-operator valued measurements (POVM) $\{A_{+|0},A_{-|0}\}$
and $\{A_{+|1},A_{-|1}\}$. The possible outcomes of each such
measurement are $+$ or $-$. Therefore, they can be assumed to be
dichotomic observables. Similarly, Bob also can measure his
subsystem chosen randomly from two $\pm$ valued POVM
$\{B_{+|0},B_{-|0}\}$ and $\{B_{+|1},B_{-|1}\}$. To explain the
non-local property, Hardy puts following three constraint on the
joint probabilities:
\begin{equation}\label{cond3}
\begin{array}{lll}
p(+,+|A_{0},B_{0})=0,\\
p(+,-|A_{1},B_{0})=0,\\
p(-,+|A_{0},B_{1})=0,
\end{array}
\end{equation}
where, $A_{i}\equiv A_{+|i}-A_{-|i}$ and $B_{i} \equiv B_{+|j}-B_{-|j}$
for $i,j\in\{0,1\}$ and $p(a,b|A_{x},B_{y})$ denotes the joint probability of observing result, with $a,b=\pm $ under local setting $x,y\in\{0,1\}$. Note that under LHV theory each of the joint probability $p(a,b|A_{x},B_{y})$ can be expressed as $p(a,b|A_{x},B_{y})=\sum_{\lambda}p_{\lambda}(a|A_{x})p_{\lambda}(b|B_{y})$. Therefore, one can easily get the following constraint
\begin{equation}\label{phardy}
p_{Hardy}\equiv p(+,+|A_{1},B_{1})=0.
\end{equation}
However, Hardy showed that there exists two-qubits non-maximal correlation which satisfies all the three conditions of (\ref{cond3}) but can give a non-zero value of $p_{Hardy}$ {\em i.e.}, $p_{Hardy}\equiv p(+,+|A_{1},B_{1})>0$. 
More precisely, it has been proved that for a given pair of
dichotomic observables on each site there exists an unique
two-qubits non-maximally entangled state which satisfies all the
three conditions of (\ref{cond3}) and violate the condition of
(\ref{phardy}) \cite{Kar97} and the maximal value of $p_{Hardy}$ can
go up to $\frac{5\sqrt{5}-11}{2}$ in the case of two qubits
preparation \cite{Jor94}. Recently, Rabelo {\it et al.} \cite{SC}
proved that the maximal Hardy probability ($p_{Hardy}$) has no
advantage in higher-dimension quantum systems, they also proposed an
non-ideal version of the paradox by introducing external noisy as
the following section.
\subsubsection{Noisy Hardy paradox} It is quite obvious that the three joint probabilities given in (\ref{cond3}) may not be identically zero in practice due to
the noise introduced by external environment and/or by imperfectness of the devices. Therefore, it is worthy to study the imperfect case along with the ideal scenario of the Hardy test. Let us denote our noisy parameter as $\epsilon$ and consider each of the concerning three joint probabilities are bounded
by $\epsilon$. Hence, the original constraints (\ref{cond3}) on the joint probabilities reduce to following inequalities
\begin{equation}\label{cond3New}
\begin{array}{lll}
p(+,+|A_{0},B_{0})\leq\epsilon,\\
p(+,-|A_{1},B_{0})\leq\epsilon,\\
p(-,+|A_{0},B_{1})\leq\epsilon.
\end{array}
\end{equation}
From the famous CH inequality \cite{CH74}, or rather its left hand
side \bea\label{CH} &-1\leq
p(+,-|A_1,B_0)+p(-,+|A_0,B_1)+p(-,-|A_0,B_0)&\nonumber\\&-p(-|A_0)-p(-|B_0)-p(+,+|A_1,B_1),&
\eea we get the following LHV bound on the maximal value of the
Hardy probability $p_{Hardy}$
\begin{equation}
\begin{array}{lll}
p_{Hardy} \\ \leq
p(+,+|A_{0},B_{0})+p(+,-|A_{1},B_{0})+p(-,+|A_{0},B_{1}) \\ \leq
3\epsilon,
\end{array}
\end{equation}
the first inequality of the above expression can be reached
immediately from (\ref{CH}), once one notices that \ben
p(+,+|A_0,B_0)= p(-,-|A_0,B_0)-p(-|A_0)-p(-|B_0) +1.\een
Note that $p_{Hardy}$ attends its algebraic maximum when the corresponding noisy parameter $\epsilon$ reaches the value $\frac{1}{3}$. Hence, it is worthy to discuss the noise Hardy test for $0\leq\epsilon <\frac{1}{3}$. If the noisy parameter
is larger than $\frac{1}{3}$, then the maximal bound of Hardy's probability $p_{Hardy}$ has no advantage over the LHV bound.
\subsection{Cabello paradox} In 2002, Cabello introduced another logical structure to prove Bell's theorem without inequality for three-particle GHZ and W states. Later Liang and Li demonstrated that the argument is also applicable for two two-level systems \cite{Liang}.
The mathematical formulation of Cabello's non-locality argument for two
two-level systems is as follows:
\begin{equation}
\begin{array}{lll}
p(+,+|A_{0},B_{0})=q_1,\\
p(+,-|A_{1},B_{0})=0,\\
p(-,+|A_{0},B_{1})=0,\\
p(+,+|A_{1},B_{1})=q_4.
\end{array}
\end{equation}
One can also check that this set of conditions cannot be satisfied by any LHV theory as long as $q_4>q_1$. Therefore, LHV bound on the probability of success of Cabello's argument $p_{Cabello}\equiv q_4-q_1$ is given by
\begin{equation*}
p_{Cabello}\leq0.
\end{equation*}
Note that the ideal Hardy's argument is a special case of the Cabello's argument with $q_1=0$. On the other hand, Cabello's argument can also be explained as a particular case of noisy Hardy paradox, where the first joint probability of (\ref{cond3}) introduce a noise $q_1$ and other two concerning joint probabilities do not introduce any noise.



In 2006, Kunkri {\it et al}. \cite{Kunkri} proved that the maximum
probability of success in Cabello's argument for two two-qubits
system is approximately $11\%$, which is larger than the original
Hardy's success probability. Here, we show that the maximal bound of
success probability of Cabello's case has also no advantage in
higher-dimension quantum systems and it can be used to generate much
more randomness compare to the original Hardy test. More
interestingly, Cabello's argument can also be transformed to the
{\em Dimension Witness (DW)} paradox by considering that Alice has
the free will assumption. Farther, this can be used to construct new SDITRNG protocol based on non-locality test without inequality.
\section{Device-Independent randomness based on Hardy's paradox}
Consider that state $\rho$ is shared between Alice and Bob. If Alice and Bob perform the measurement $(A_x,B_y)$ on their respective part of $\rho$, then the joint probability of getting the outcome $(a,b)$ is 
\begin{equation*}
p(a,b|A_{x},B_{y})=Tr(\rho A_{a|x}\otimes B_{b|y}).
\end{equation*}
 Note that there is no constraint on the dimension of the system. Therefore, without loss of generality, we assume that $\rho$ is pure and all the concerning measurements are projective. It is obvious that the measurement outcomes must obey the causality principle {\em i.e.}, the corresponding joint probabilities satisfy the following non-signalling conditions
\begin{equation}\label{ns}
\begin{array}{lll}
p(a|A_{x})=\sum_{b}p(a,b|A_{x},B_{0})=\sum_{b}p(a,b|A_{x},B_{1}),\\
p(b|B_{y})=\sum_{a}p(a,b|A_{0},B_{y})=\sum_{a}p(a,b|A_{1},B_{y}),
\end{array}
\end{equation}
where $p(a|A_x)$ and $p(b|B_y)$ denote the marginal probability. In
the original DITRNG protocol, non-local correlations are used to
certify the presence of genuine randomness in quantum theory. Note
that the randomness of measurement outcome is sovereign and it
independent of entanglement. In this context, it has been proved
that non-locality with tiny amount of entanglement can generate full
randomness asymptotically \cite{acin}. The parameter that used to
estimate randomness of the measurement outcomes $a$ and $b$
conditioned on the input values $x$ and $y$ is the min-entropy
function \cite{renner1,renner2}. Since the state preparation and the
measurement have no restriction in device independent protocol,
Alice and Bob's measurement equipments can be
assumed to be black boxes. Let $x\in\{0,1\}$ and $y\in\{0,1\}$ be the inputs of Alice and Bob's measurement black box respectively, 
then the min-entropy function for the given inputs $(x,y)$ can be expressed as
\begin{equation}
H_{\infty}(a,b|x,y)\equiv-\log_{2}[\max_{\{a,b\}}{p(a,b|A_{x},B_{y})}].
\end{equation}
where $\max_{\{a,b\}}p(a,b|A_{x},B_{y})$ denotes the maximal probability over all possible outcomes $(a,b)$. The min-entropy can reach the maximal value if
$p(a,b|A_{x},B_{y})=\frac{1}{4}$ for all $a$ and $b$.

Presently, the {\em Semi-Definite Programming (SDP)} is a prominent
method to solve the convex optimization problem, which concerned
with linear objective function over the semi-definite matrices.
Follow the method suggested by Navascues {\it et al.} \cite{sdp2},
we solve the following maximal guessing probability optimization
problem with SDP method
\begin{equation}
\begin{array}{lll}
\text{Minimize}&:& \max_{\{a,b\}} p(a,b|A_{0},B_{0})\\
\text{Subject to}&:& \Delta_{Hardy}\geq0,\\
&&p(+,+|A_{0},B_{0})=0,\\
&&p(+,-|A_{1},B_{0})=0,\\
&&p(-,+|A_{0},B_{1})=0,\\
&&p(+,+|A_{1},B_{1})=p_{Hardy},
\end{array}
\end{equation}
where $\Delta_{Hardy}=[\Delta_{ij}]$ is the positive semi-definite matrix with $ \Delta_{ij}=Tr(E_i^{\dag}E_j\rho)$ for $E_i,E_j\in\{I, A_{a|x}, B_{b|y}, A_{a|x}B_{b|y}\}$. Note that these
measurement operators should also satisfy the hermiticity
($A_{a|x}=A_{a|x}^{\dag}, B_{b|y}=B_{b|y}^{\dag}$),
orthogonality ($A_{a|x}A_{a'|x}=\delta_{aa'}A_{a|x},
B_{b|y}B_{b'|y}=\delta_{bb'}B_{b|y}$),
completeness ($\sum_a A_{a|x}=\sum_b B_{b|y}=1 $),
commutativity ($[A_{a|x}, B_{b|y}]=0$) and the no-signaling conditions (\ref{ns}). By
applying the SDP method and the Sedumi toolbox \cite{sdp}, we
calculate the maximal min-entropy bound
$H_{\infty}(a,b|A_{0},B_{0})$ with different Hardy parameter
$p_{Hardy}$.
\begin{figure}[!h]\center
\resizebox{9cm}{!}{
\includegraphics{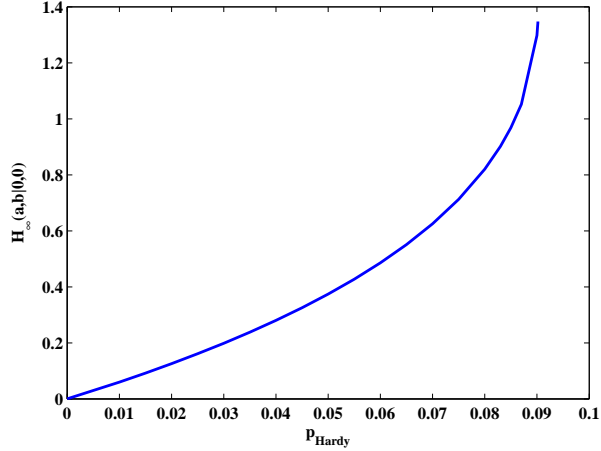}}
\caption{Maximal min-entropy bound $H_{\infty}(a,b|A_{0},B_{0})$
with different Hardy parameter $p_{Hardy}$}\label{fig1}
\end{figure}
Calculation result shows that the maximal randomness can reach up to
1.35 if the corresponding Hardy's probability attends its maximal
value $\frac{5\sqrt{5}-11}{2}$. Comparing with the CHSH inequality
based DITRNG protocol, Hardy paradox generates much more true random
number in optimal case.

By applying similar SDP optimization method, we can estimate the
generated randomness based on Noisy Hardy paradox
\begin{equation}
\begin{array}{lcl}
\text{Minimize}&:& \max_{\{a,b\}} p(a,b|A_{0},B_{0})\\
\text{Subject to}&:& \Delta_{Noisy~Hardy}\geq0,\\
&&p(+,+|A_{0},B_{0})\leq\epsilon,\\
&&p(+,-|A_{1},B_{0})\leq\epsilon,\\
&&p(-,+|A_{0},B_{1})\leq\epsilon,\\
&&p(+,+|A_{1},B_{1})=\delta_{\epsilon}.
\end{array}
\end{equation}
where $\Delta_{Noisy~Hardy}$ is the positive semi-definite matrix,
$\epsilon$ is the noisy Hardy parameter, $\delta_{\epsilon}$ is the
maximal joint probability $p(+,+|A_{1},B_{1})$ by considering the constraint on three joint probabilities, which will reduce to
$\frac{5\sqrt{5}-11}{2}$ when the noisy Hardy parameter reduces to 0.
The noisy Hardy paradox can show contradictions with LR theory only when $\delta_{\epsilon}$ is larger than $3\epsilon$ i.e., when it's violate the associate LHV bound. We apply this non-local property to
estimate randomness of the measurement outcomes, the maximal
min-entropy bound $H_{\infty}(a,b|A_{0},B_{0})$ with different noisy
Hardy parameter $\epsilon$ is given in Fig.2.
\begin{figure}[!h]\center
\resizebox{9cm}{!}{
\includegraphics{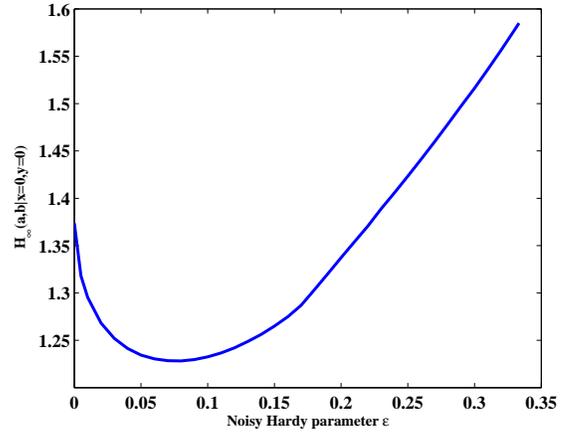}}
\caption{Maximal min-entropy bound $H_{\infty}(a,b|A_{0},B_{0})$
with different noisy Hardy parameter $\epsilon$}
\end{figure}
From the calculation result, we find that the maximal randomness can
reach 1.58 when the associated noisy Hardy parameter reaches its
maximal value 0.3333. Correspondingly, the maximal quantum value can
reach 0.99995, which is lager than the LHV bound 0.99990. Comparing
with the CHSH inequality and original Hardy paradox, the noisy Hardy
paradox can generate much more random number when
$\delta_{\epsilon}$ reach the maximal value correspondingly.

Randomness of the DITRNG protocol based on Cabello paradox can also
be analyzed with the similar method,
\begin{equation}
\begin{array}{lcl}
\text{Minimize}&:& \max_{\{a,b\}} p(a,b|A_{0},B_{0})\\
\text{Subject to}&:& \Delta_{Cabello}\geq0,\\
&&p(+,-|A_{1},B_{0})=0,\\
&&p(-,+|A_{0},B_{1})=0,\\
&&p(+,+|A_{1},B_{1})-p(+,+|A_{0},B_{0})=p_{Cabello}.
\end{array}
\end{equation}
where $\Delta_{Cabello}$ is positive semi-definite matrix,
$p_{Cabello}$ is the success prability of Cabello paradox, which is the joint
probability deviation value $p(+,+|A_{1},B_{1})-p(+,+|A_{0},B_{0})$.
The SDP calculation result shows that maximal value of $p_{Cabello}$
is 0.10784 in arbitrary high dimension quantum system, which prove
that high dimension system has no advantage to improve
$p_{Cabello}$. The Cabello paradox has the non-local property when
$p_{Cabello}$ is larger than zero, in which case the protocol can
generate the true random number, the maximal min-entropy bound of
$H_{\infty}(a,b|A_{0},B_{0})$ with different Cabello paradox
parameter $p_{Cabello}$ can be given in Fig.3.
\begin{figure}[!h]\center
\resizebox{9cm}{!}{
\includegraphics{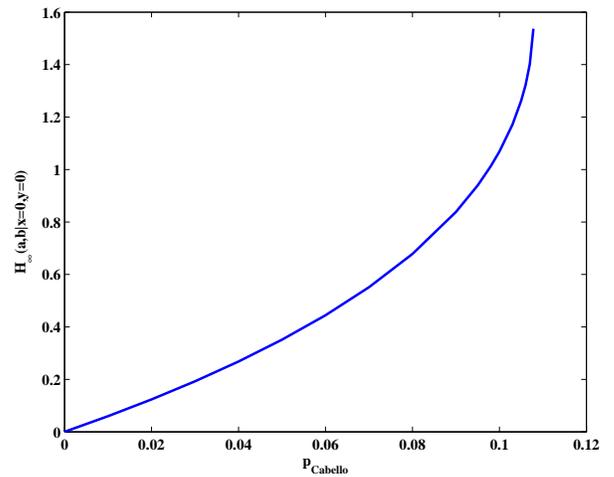}}
\caption{ Maximal min-entropy bound of $H_{\infty}(a,b|A_{0},B_{0})$
with different Cabello paradox parameter $p_{Cabello}$}
\end{figure}
Calculation shows that the maximal randomness can reach up to value
1.56 and the maximum attends when the associate Cabello's parameter
reaches its maximal value 0.10784. Thus, the efficiency of
generating randomness is higher compare to the original Hardy
paradox in optimal case. More importantly, the Cabello paradox can
be transformed to a dimension witness paradox (from entanglement
based protocol to state preparation and measurement based protocol),
which can be used for constructing one way SDITRNG protocol.
\section{ Semi-Device-Independent randomness based on Hardy inequality}
In a recent work by Li {\em et al.} \cite{Li3} proved that CHSH
inequality can be transformed to a dimension witness inequality by
considering Alice has random measurement outcomes with arbitrary
input random number (which is equal to free will assumption). By
applying the similar technique, we present the first dimension
witness paradox without inequality.
\begin{equation}
\begin{array}{lll}
p(+,+|A_{0},B_{0})=q_1,\\
p(+,-|A_{1},B_{0})=0,\\
p(-,+|A_{0},B_{1})=0,\\
p(a|A_{x})=p(a|A_{x},B_{y})=\frac{1}{2},\\
p(+,+|A_{1},B_{1})=q_4.\\
\end{array}
\end{equation}
Similar to the LHV theory, the classical mechanics theory has the
Cabello paradox parameter restriction $p_{Cabello}\leq0$, while
$p_{Cabello}>0$ guarantee that
 the system can only be explained by the quantum mechanics.

Since the dimension witness can be used to generate {\em semi-device independent true random
number}, where the randomness has no more restriction about the state
preparation and measurement except the fact that the dimension of Hilbert space associated to each subsystem is two. Then we apply the SDP method to
calculate the maximal guessing probability
$\max_{a,b}p(b|A_{x},a,B_{y})$ by considering the quantum dimension
witness based on Cabello paradox.
\begin{equation}
\begin{array}{lcl}
\text{Minimize}&:& \max_{\{a,b\}} p(b|A_{0},a,B_{0})\\
\text{Subject to}&:& \Delta_{Cabello}\geq0,\\
&&p(+,-|A_{1},B_{0})=0,\\
&&p(-,+|A_{0},B_{1})=0,\\
&&p(a|A_{x})=p(a|A_{x},B_{y})=\frac{1}{2},\\
&&p(+,+|A_{1},B_{1})-p(+,+|A_{0},B_{0})=p_{Cabello},
\end{array}
\end{equation}
where the Cabello paradox parameter $p_{Cabello}$ can reach the
maximal value 0.08279, which is smaller than the original Cabello
paradox parameter, the reason for which is that the original Cabello
paradox can not get full random measurement outcome in one side
black box. Since have proved that SDP method can also be used in
SDITRNG protocol \cite{Li3}, the maximal min-entropy bound of
$H_{\infty}(b|A_{0},a,B_{0})$ with different Cabello paradox
parameter $p_{Cabello}$ is given in Fig.4.
\begin{figure}[!h]\center
\resizebox{9cm}{!}{
\includegraphics{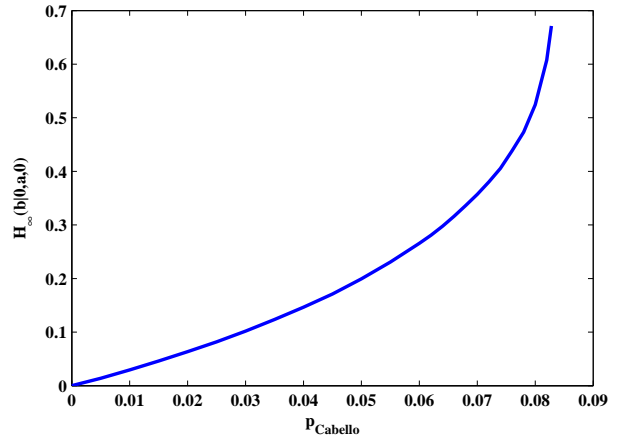}}
\caption{ Maximal min-entropy bound of $H_{\infty}(b|A_{0},a,B_{0})$
with different Cabello paradox parameter $p_{Cabello}$}
\end{figure}
From the calculation result, we find that the maximal randomness can
reach 0.68 when the Cabello's parameter reaches the maximal value.
Thus efficiency of this new protocol is evidently larger than the
original SDITRNG protocol, which can generate maximum 0.23 random
number and the maximum value attends when the dimension witness
inequality value is 2.828.
\section{ Conclusion}
In conclusion, We have proposed {\em device
independent true random number generation} protocols based on Hardy paradox, Noisy Hardy
paradox and Cabello paradox respectively. All of these protocols can
generate much more true random numbers compare to all other DITRNG protocols based on some Bell-CHSH type inequalities. More interestingly,
we also proposed a new dimension witness paradox by using the Cabello
argument and consequently developed a {\em semi-device independent true random number generation} protocol based on this dimension witness paradox. The subject of {\em device independent (DI)} proof for various quantum protocols remains a complicated area till date and knowledge in this regards is day by day increasing by developing new proofs for DI protocols. In this regard, a new kind of DI proof for quantum key distribution based on Hardy's paradox has been proposed very recently \cite{qkd}. It is an open problem to discuss other quantum information protocols
based on non-inequality paradoxes.
\section{Acknowledgements}
H-W.L., G-C.G. and Z-F.H. are supported by the
the National Natural Science Foundation of China (Grant Nos.
61101137, 61201239, 61205118 and 11304397). M.P. is supported by UK
EPSRC.

\end{document}